\documentclass[12pt,letterpaper]{article} 

\usepackage[utf8]{inputenc}
\usepackage[T1]{fontenc}
\usepackage{times} 
\usepackage{amsmath,amsfonts,amssymb}
\usepackage{graphicx}
\usepackage{setspace}
\usepackage[left=0.875in, right=0.875in, top=1.0in, bottom=1.25in]{geometry}

\usepackage{titlesec}
\titleformat{\section}{\normalfont\large\bfseries\uppercase}{\thesection.}{0.5em}{}
\titleformat{\subsection}{\normalfont\large\bfseries}{\thesubsection}{0.5em}{} %

\usepackage{amsmath,amsfonts,amssymb}
\usepackage{graphicx}
\usepackage[colorlinks=true, allcolors=blue]{hyperref}
\usepackage{authblk} 

\usepackage{fancyhdr}
\date{\today}        
\date{Octobre 2026}  

\title{BRICS Southern Observatory (BRICS-SO) Project: Invitation to African countries to invest in Chile}

\author[a,b]{Massinissa Hadjara}
\author[c]{Amokrane Berdja}
\author[d]{Swagat R. Das}
\author[e]{Elisson S. G. de Almeida}
\author[f]{Slobodan Jankov}
\author[g]{Vladmir Lukin}
\author[h]{Abdelkarim Boskri}
\author[h]{Zouhair Benkhaldoun}
\author[i]{Jinghui Zheng}
\affil[a]{National Institute of Astronomical Optics \& Technology, Chinese Academy of Sciences, Nanjing 210042}
\affil[b]{China Chinese High Angular Resolution Southern Astronomical Laboratory (CHARSAL)/Astro-Photonics Laboratory/Space and Planetary Exploration Laboratory (SPEL), Dept. of Electrical Engineering, FCFM, Universidad de Chile, Av. Tupper 2007, Santiago, Chile}
\affil[c]{Centro de Astronomıa (CITEVA), Universidad de Antofagasta, Avenida Angamos 601, Antofagasta 1270300, Chile}
\affil[d]{Departamento de Astronomıa, Universidad de Chile, Casilla 36-D, Santiago, Chile}
\affil[e]{Instituto de Fisica y Astronomia, Facultad de Ciencias, Universidad de Valparaıso, Av. Gran Bretana 1111, Valparaiso, Chile}
\affil[f]{Astronomical Observatory, Volgina 7, PO Box 74 11060 Belgrade, Serbia}
\affil[g]{Zuev Institute of Atmospheric Optics, Siberian Branch of the Russian Academy of Sciences, pl. Akademika Zueva 1, 634055 Tomsk, Russian Federation}
\affil[h]{Oukaimeden Observatory, Laboratory of High Energy Physics and Astrophysics, FSSM, Cadi Ayyad University, Marrakech Morocco}
\affil[i]{Shanghai Astronomical Observatory, CAS, 80 Nandan Road, Shanghai 200030, China}

\begin{document} 
\maketitle

\thispagestyle{fancy} 
\pagestyle{fancy}     

\begin{abstract}
Recently, the BRICS countries created a promising astronomy consortium (\url{https://www.bricsastronomy.org/}), which was joined last year by some emerging countries, including two African countries, namely: Egypt and Ethiopia. Certainly, other African countries will join BRICS+ in the years to come. Thus, by pooling their efforts, emerging countries (BRICS+) are sure to quickly catch up with their technological delay in the field of astronomy, particularly through ambitious and large-scale projects.
Chile, due to the exceptional climatic conditions of its North, is recognized as the world capital of astronomy where all the largest consortia have settled there and applied all their know-how.
The official launch of BRICS (and then its expansion to other countries; BRICS+) is welcomed as an excellent opportunity to think big in the development of astronomy. China has shown great determination in being the first country in the South to establish astronomical facilities in Chile, thus opening a possible opportunity for BRICS+ astronomers.

Ambitious astronomical instrumentation is essential for the future of Astronomy and Astrophysics. Access to existing facilities is becoming more and more difficult because of saturation. Existing programs like the VLT-I/ELT and their instruments necessitated decades of collaboration between international institutions. It is therefore necessary for the BRICS+ astronomical community to collaborate together on ambitious projects without wasting time in order to offer a new pole with new opportunities for the researchers. This pole (BRICS-SO), in addition to attracting highly qualified people, will also help train new generations in all our countries, and encourage exchanges and collaborations.
\end{abstract}


\section{Introduction}

\begin{figure}
\centering
  \includegraphics[scale=.4]{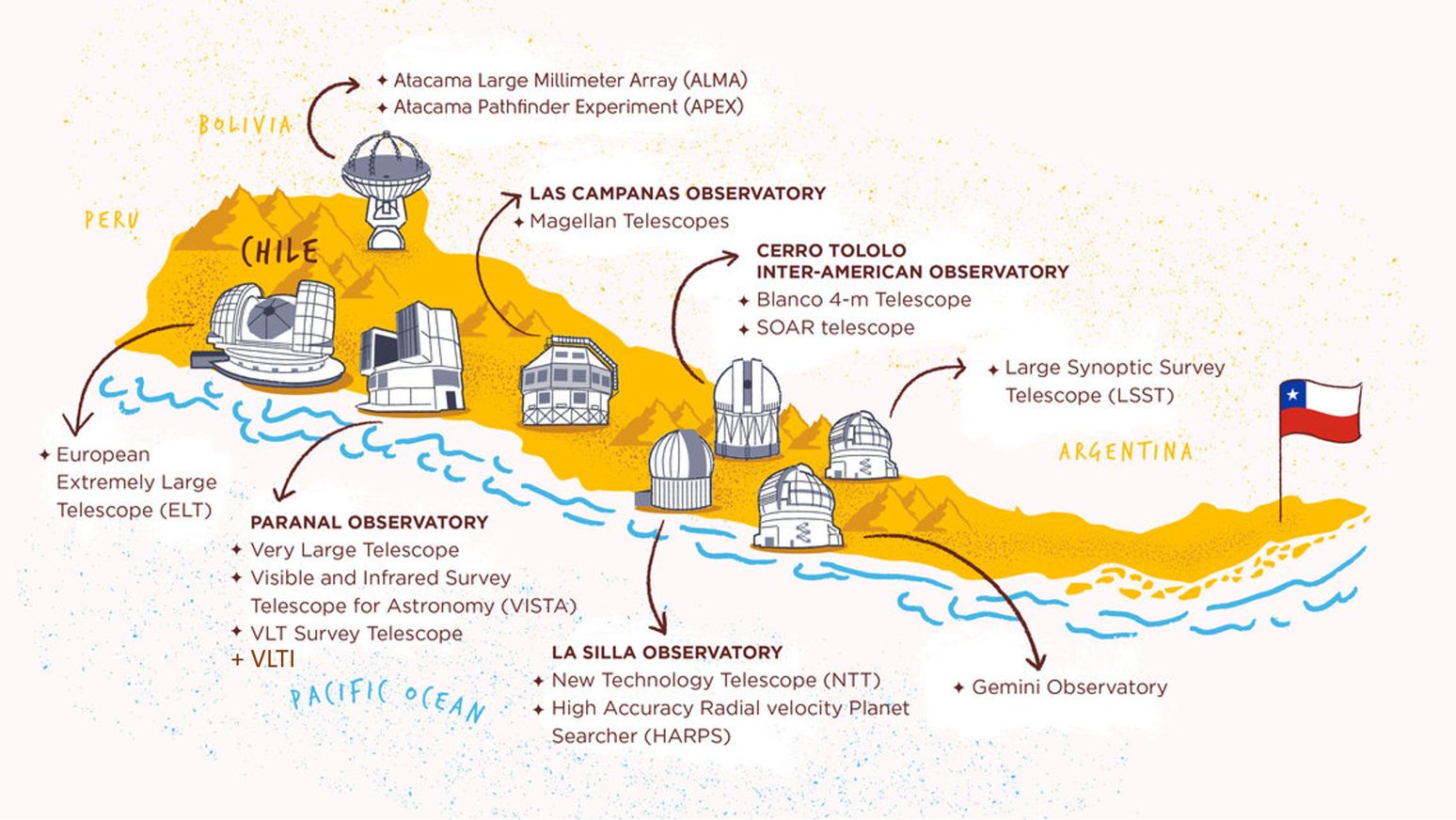}
  \caption{70\% of all the world's astronomical observation capacity are located in northern Chile (\textcircled{c}Artwork by Sandbox Studio, Chicago with Pedro Rivas).}
  \label{Fig1}
\end{figure}

Hosting more than 70\% of all the world's astronomical observation capacity (as shown in figure \ref{Fig1}), Chile can be considered the world capital of astronomy. In fact, the Chilean north offers exceptional observation conditions and the Chilean government's policy grants foreign observatories every facility to establish themselves on its territory. Several countries have already established themselves there in exchange for 10\% of observations for the benefit of Chilean astronomers: Europe via the European Southern Observatory (ESO) with the VLTI or soon the ELT, the United States of America via the National Optical Astronomy Observatory (NOAO) at Cerro Tololo and associated, Japan via ALMA (an international partnership between Europe, North America and East Asia), Russia via small observatories/telescopes, and more recently China.
On the other hand, no African country is represented there. Only South Africa participates in the Cherenkov Telescope Array South project.

\section{Optical Interferometry}

\begin{figure}
\centering
  \includegraphics[scale=.4]{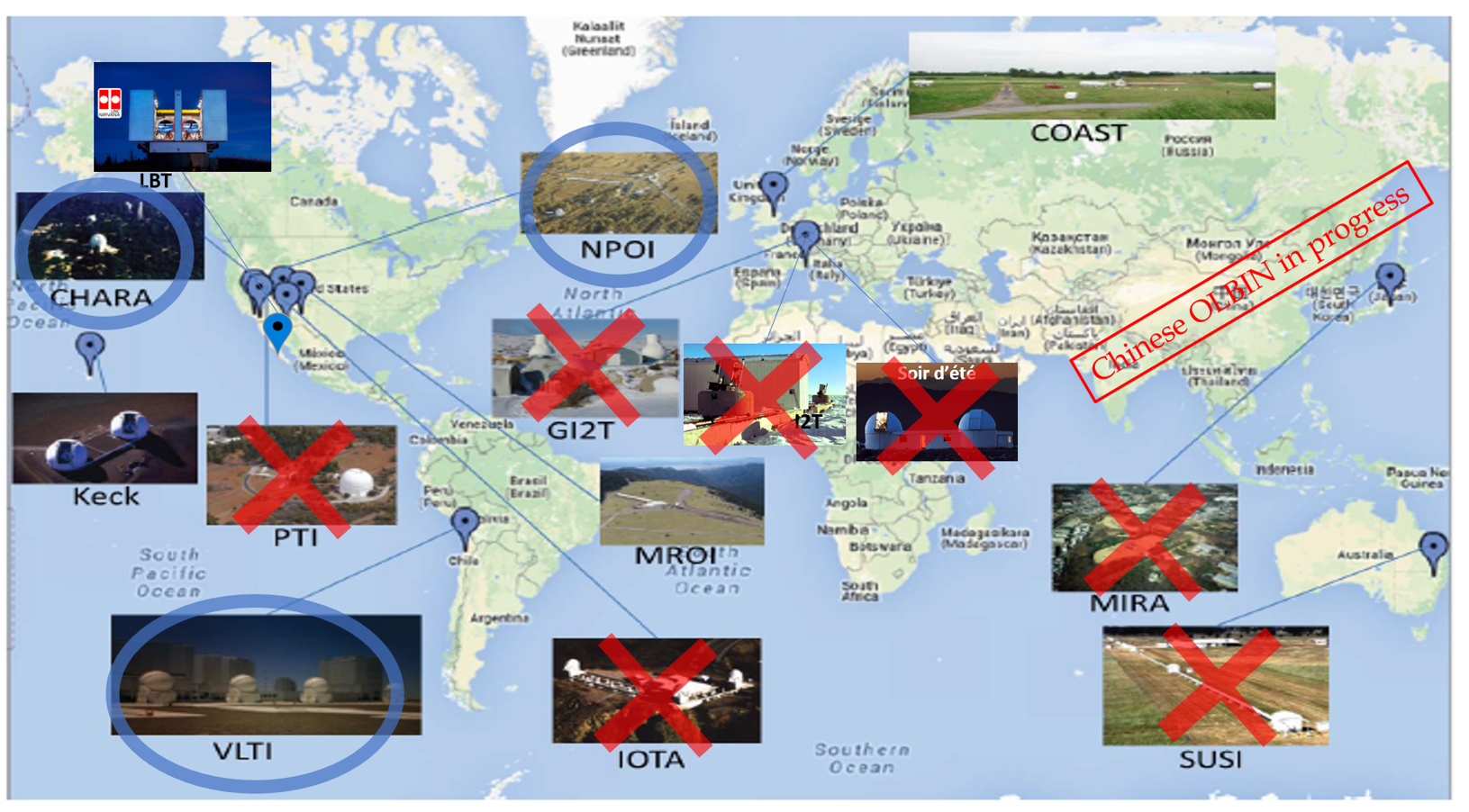}
  \caption{Interferometers around the world. Those circled in blue are those that are capable of providing closure phase (i.e. capable of combining light from 3 or more telescopes). Of all the emerging countries, only China is currently developing its own 3-telescope interferometer in China \cite{2024SPIE13095E..24X}.}
  \label{Fig2}
\end{figure}

\begin{figure}
\centering
  \includegraphics[scale=.4]{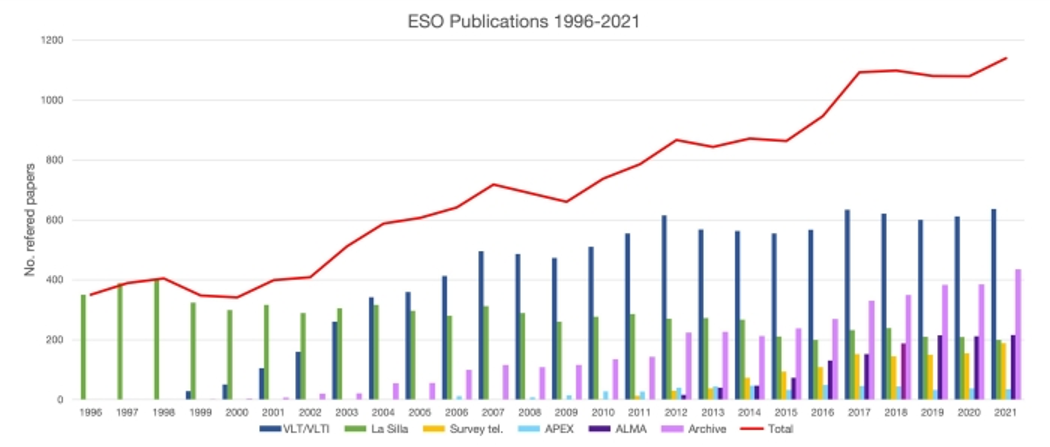}
  \caption{Publications of the European Southern Observatory (ESO) from 1996 to 2021 (\url{https://www.eso.org/}). We can clearly appreciate the increasing contribution of interferometry (via the VLTI) over the years.}
  \label{Fig3}
\end{figure}

The construction of large optical/IR single aperture telescopes has a technical limit. Nowadays, the maximum aperture is 39 meters for ESO's Extremely Large Telescope (ELT; \url{https://elt.eso.org}).
A practical solution is to use the technique of interferometry, which involves combining the light from several small and medium-sized telescopes over large distances (so-called baselines) of several meters. We certainly lose, in fact, in terms of flux but we gain enormously in terms of angular resolution.
The exact name of this technique, which appeared in the 1970s, is Optical Long Baseline Interferometry (OLBIn) \cite{1975ApJ...196L..71L}, although interferometry instruments, using aperture masks and mirror arms with more limited baselines (a few meters), have been known since the 19th century \cite{1999AAS...195.2301L}.
Since then, the OLBIn has experienced a global boom, as shown in figure \ref{Fig2}. Currently, only a few interferometers are still operational, mainly in North America. But the most sophisticated and productive IR interferometer remains the Very Large Telescope Interferometer (VLTI) in the Atacama Desert (Chile), as shown in figure \ref{Fig3}.

ESO succeeded in building the VLTI, after a long (almost 25 years) and complicated struggle, both technical and collaborative between several European institutions:
\begin{itemize}
\item December 1977: The VLTI was suggested by Prof. Antoine Labeyrie at the ESO conference in Geneva (Switzerland),
\item 1983 : The decision to build the VLT was made, 
\item 1987: It was agreed to include an interferometric component (VLTI),
\item 1990: ESO chose to install the VLT/VLTI in the north of Chile,
\item 1991-1998: Construction of the VLTI (7 years),
\item March 17, 2001 (almost 23 years ago): The “first fringes” with VLTI/VINCI
\end{itemize}
From the first suggestion (1977) to the first light (2001), 24 years had passed (47 years ago)

Meanwhile, Africa and the BRICS+ Astronomy Consortium (BAC) in general, unfortunately, are not beginning to think about using this technology (or at least preparing the ground for the future).

The VLTI is by far the most productive optical interferometer in the world. Combining the highest sensitivity thanks to its 4 Unit Telescopes (UTs) and a good imaging capability boosted by its 4 Auxiliary Telescopes (Ats) that can be moved around to improve the (u,v)-coverage, its resolution is of a few milliarcseconds for imaging and can be pulled below 10 micro-arcsecond by Spectro-Astrometry from color Phase Differential Interferometry (PDI). Its first generation instruments (AMBER in the H and K bands and MIDI in the N band) have provided major breakthroughs in stellar physics (the envelopes of Be, B[e] and Herbig stars \cite{2007AA...464...73M,2009AA...504..915C,2008ApJ...676..490K}, the MOLsphere of giant stars \cite{2011AA...529A.163O,2017Natur.548..310O,2019MNRAS.489.2595H}, LBVs \cite{2007AA...464...87W,2016AA...594A.106W}, fundamental stellar parameters \cite{2009AIPC.1094..792D}, discs in YSO[16] and many other topics, including the structure of dust torus in AGNs \cite{2004AA...418..177V,2007AA...474..837T,2013Msngr.154...62B,2018Natur.563..657G,2020NatAs...4..517W,2022Natur.602..403G}, thanks to the 2nd generation instruments; GRAVITY and MATISSE.
The new challenge of OLBIn is to observe and study extragalactic objects. In particular Active Galactic Nuclei (AGN); thanks to the Spectro-Astrometry of the Board Line Regions (BLR) of Seyfert 1 (Sy1) related to synchrotron radiation. Thus, we can directly infer the distances and masses of AGNs both by Spectro-Astrometry (SA), which provides the angular sizes using Phase Differential Interferometry (PDI) and by Reverberation-Mapping (RM), which provides the linear sizes, using single-telescope spectroscopy (SARM[20]). Indeed, the direct estimation of distances in an unprecedented range of cosmic distances has a potentially decisive impact on the Hubble tension problem which is currently one of the major questions of fundamental astrophysics. Currently, OLBIn can observe objects of magnitude K=13, thanks to the Fringe Tracker (FT) of VLTI/GRAVITY. but thanks to an ambitious new generation of FTs, e.g. hierarchical FT (HFT) \cite{2021MNRAS.506.1364B,2024SPIE...13095..89P,2024SPIE...13095..90A}, it has been proven that we can reach K=16 (increasing by the way the number of possible AGN observations by a factor of 100 in the whole IR range, where currently with FT-GRAVITY we can only observe a few).

\section{The Dream}

See African countries and all emerging countries in general (via BAC) pool their efforts and start thinking about future interferometric projects in Chile and Antarctica.

\section{Discussions \& Conclusion}

We must all together start working on such large-scale and ambitious future projects, now. It is our sacred duty as scientists towards future generations\footnote{The tipping point for changing/improving things in social communities is between 21\% and 25\% \cite{2018Sci...360..6393}}.

\section*{Acknowledgement}

E.S.G. de Almeida has been financially supported by ANID Fondecyt postdoctoral grant folio N. 3220776.

\bibliography{report} 

@article{1975ApJ...196L..71L,
       author = {{Labeyrie}, A.},
        title = "{Interference fringes obtained on Vega with two optical telescopes.}",
      journal = {\text{Astrophys. J. Lett.}},
         year = 1975,
        month = mar,
       volume = {196},
        pages = {L71-L75},
          doi = {10.1086/181747},
       adsurl = {https://ui.adsabs.harvard.edu/abs/1975ApJ...196L..71L}
}

@INPROCEEDINGS{1999AAS...195.2301L,
       author = {{Lawson}, P.~R.},
        title = "{A History of Stellar Interferometry: from 1868 to the present}",
    booktitle = {American Astronomical Society Meeting Abstracts},
         year = 1999,
       series = {American Astronomical Society Meeting Abstracts},
       volume = {195},
        month = dec,
          eid = {23.01},
        pages = {23.01},
       adsurl = {https://ui.adsabs.harvard.edu/abs/1999AAS...195.2301L}
}

@INPROCEEDINGS{2024SPIE13095E..24X,
       author = {{Xu}, Teng and {Hou}, Yonghui and {Jiang}, Fanghua and {Zhou}, Guohua and {Sun}, Yue and {Wu}, Xuhao and {Li}, Xiaofei and {Sun}, Hemin and {Wei}, Wei and {Hadjara}, Massinissa and {Hu}, Zhongwen},
        title = "{Progress of the 100m-baseline optical interferometer in China}",
    booktitle = {Optical and Infrared Interferometry and Imaging IX},
         year = 2024,
       editor = {{Kammerer}, Jens and {Sallum}, Stephanie and {Sanchez-Bermudez}, Joel},
       series = {Society of Photo-Optical Instrumentation Engineers (SPIE) Conference Series},
       volume = {13095},
        month = aug,
          eid = {1309524},
        pages = {1309524},
          doi = {10.1117/12.3018797},
       adsurl = {https://ui.adsabs.harvard.edu/abs/2024SPIE13095E..24X}
}

@article{2007AA...464...73M,
       author = {{Meilland}, A. and {Stee}, P. and {Vannier}, M. and {Millour}, F. and {Domiciano de Souza}, A. and {Malbet}, F. and {Martayan}, C. and {Paresce}, F. and {Petrov}, R.~G. and {Richichi}, A. and {Spang}, A.},
        title = "{First direct detection of a Keplerian rotating disk around the Be star {\ensuremath{\alpha}} Arae using AMBER/VLTI}",
      journal = {\text{Astron. Astrophys.}},
         year = 2007,
        month = mar,
       volume = {464},
       number = {1},
        pages = {59-71},
          doi = {10.1051/0004-6361:20064848},
archivePrefix = {arXiv},
       eprint = {astro-ph/0606404},
 primaryClass = {astro-ph},
       adsurl = {https://ui.adsabs.harvard.edu/abs/2007A&A...464...59M}
}

@article{2009AA...504..915C,
       author = {{Carciofi}, A.~C. and {Okazaki}, A.~T. and {Le Bouquin}, J.-B. and {{\v{S}}tefl}, S. and {Rivinius}, Th. and {Baade}, D. and {Bjorkman}, J.~E. and {Hummel}, C.~A.},
        title = "{Cyclic variability of the circumstellar disk of the Be star {\ensuremath{\zeta}} Tauri. II. Testing the 2D global disk oscillation model}",
      journal = {\text{Astron. Astrophys.}},
         year = 2009,
        month = sep,
       volume = {504},
       number = {3},
        pages = {915-927},
          doi = {10.1051/0004-6361/200810962},
archivePrefix = {arXiv},
       eprint = {0901.1098},
 primaryClass = {astro-ph.SR},
       adsurl = {https://ui.adsabs.harvard.edu/abs/2009A&A...504..915C}
}

@article{2008ApJ...676..490K,
       author = {{Kraus}, Stefan and {Preibisch}, Thomas and {Ohnaka}, Keiichi},
        title = "{Detection of an Inner Gaseous Component in a Herbig Be Star Accretion Disk: Near- and Mid-Infrared Spectrointerferometry and Radiative Transfer modeling of MWC 147}",
      journal = {\text{Astrophys. J.}},
         year = 2008,
        month = mar,
       volume = {676},
       number = {1},
        pages = {490-508},
          doi = {10.1086/527427},
archivePrefix = {arXiv},
       eprint = {0711.4988},
 primaryClass = {astro-ph},
       adsurl = {https://ui.adsabs.harvard.edu/abs/2008ApJ...676..490K}
}

@article{2011AA...529A.163O,
       author = {{Ohnaka}, K. and {Weigelt}, G. and {Millour}, F. and {Hofmann}, K.-H. and {Driebe}, T. and {Schertl}, D. and {Chelli}, A. and {Massi}, F. and {Petrov}, R. and {Stee}, Ph.},
        title = "{Imaging the dynamical atmosphere of the red supergiant Betelgeuse in the CO first overtone lines with VLTI/AMBER}",
      journal = {\text{Astron. Astrophys.}},
         year = 2011,
        month = may,
       volume = {529},
          eid = {A163},
        pages = {A163},
          doi = {10.1051/0004-6361/201016279},
archivePrefix = {arXiv},
       eprint = {1104.0958},
 primaryClass = {astro-ph.SR},
       adsurl = {https://ui.adsabs.harvard.edu/abs/2011A&A...529A.163O}
}

@article{2017Natur.548..310O,
       author = {{Ohnaka}, K. and {Weigelt}, G. and {Hofmann}, K.-H.},
        title = "{Vigorous atmospheric motion in the red supergiant star Antares}",
      journal = {\text{Nature}},
         year = 2017,
        month = aug,
       volume = {548},
       number = {7667},
        pages = {310-312},
          doi = {10.1038/nature23445},
archivePrefix = {arXiv},
       eprint = {1708.06372},
 primaryClass = {astro-ph.SR},
       adsurl = {https://ui.adsabs.harvard.edu/abs/2017Natur.548..310O}
}

@article{2019MNRAS.489.2595H,
       author = {{Hadjara}, M. and {Cruzal{\`e}bes}, P. and {Nitschelm}, C. and {Chen}, X. and {Michael}, E.~A. and {Moreno}, E.},
        title = "{A CO-multilayer outer atmosphere for eight evolved stars revealed with VLTI/AMBER}",
      journal = {\text{Mon. Not. R. Astron. Soc.}},
         year = 2019,
        month = oct,
       volume = {489},
       number = {2},
        pages = {2595-2614},
          doi = {10.1093/mnras/stz2240},
archivePrefix = {arXiv},
       eprint = {1909.03997},
 primaryClass = {astro-ph.SR},
       adsurl = {https://ui.adsabs.harvard.edu/abs/2019MNRAS.489.2595H}
}

@article{2007AA...464...87W,
       author = {{Weigelt}, G. and {Kraus}, S. and {Driebe}, T. and {Petrov}, R.~G. and {Hofmann}, K.-H. and {Millour}, F. and {Chesneau}, O. and {Schertl}, D. and {Malbet}, F. and {Hillier}, J.~D. and {Gull}, T. and {Davidson}, K. and {Domiciano de Souza}, A. and {Antonelli}, P. and {Beckmann}, U. and {Bresson}, Y. and {Chelli}, A. and {Dugu{\'e}}, M. and {Duvert}, G. and {Gennari}, S. and {Gl{\"u}ck}, L. and {Kern}, P. and {Lagarde}, S. and {Le Coarer}, E. and {Lisi}, F. and {Perraut}, K. and {Puget}, P. and {Rantakyr{\"o}}, F. and {Robbe-Dubois}, S. and {Roussel}, A. and {Tatulli}, E. and {Zins}, G. and {Accardo}, M. and {Acke}, B. and {Agabi}, K. and {Altariba}, E. and {Arezki}, B. and {Aristidi}, E. and {Baffa}, C. and {Behrend}, J. and {Bl{\"o}cker}, T. and {Bonhomme}, S. and {Busoni}, S. and {Cassaing}, F. and {Clausse}, J.-M. and {Colin}, J. and {Connot}, C. and {Delboulb{\'e}}, A. and {Feautrier}, P. and {Ferruzzi}, D. and {Forveille}, T. and {Fossat}, E. and {Foy}, R. and {Fraix-Burnet}, D. and {Gallardo}, A. and {Giani}, E. and {Gil}, C. and {Glentzlin}, A. and {Heiden}, M. and {Heininger}, M. and {Hernandez Utrera}, O. and {Kamm}, D. and {Kiekebusch}, M. and {Le Contel}, D. and {Le Contel}, J.-M. and {Lesourd}, T. and {Lopez}, B. and {Lopez}, M. and {Magnard}, Y. and {Marconi}, A. and {Mars}, G. and {Martinot-Lagarde}, G. and {Mathias}, P. and {M{\`e}ge}, P. and {Monin}, J.-L. and {Mouillet}, D. and {Mourard}, D. and {Nussbaum}, E. and {Ohnaka}, K. and {Pacheco}, J. and {Perrier}, C. and {Rabbia}, Y. and {Rebattu}, S. and {Reynaud}, F. and {Richichi}, A. and {Robini}, A. and {Sacchettini}, M. and {Sch{\"o}ller}, M. and {Solscheid}, W. and {Spang}, A. and {Stee}, P. and {Stefanini}, P. and {Tallon}, M. and {Tallon-Bosc}, I. and {Tasso}, D. and {Testi}, L. and {Vakili}, F. and {von der L{\"u}he}, O. and {Valtier}, J.-C. and {Vannier}, M. and {Ventura}, N. and {Weis}, K. and {Wittkowski}, M.},
        title = "{Near-infrared interferometry of \ensuremath{\eta} Carinae with spectral resolutions of 1 500 and 12 000 using AMBER/VLTI}",
      journal = {\text{Astron. Astrophys.}},
         year = 2007,
        month = mar,
       volume = {464},
       number = {1},
        pages = {87-106},
          doi = {10.1051/0004-6361:20065577},
archivePrefix = {arXiv},
       eprint = {astro-ph/0609715},
 primaryClass = {astro-ph},
       adsurl = {https://ui.adsabs.harvard.edu/abs/2007A&A...464...87W}
}

@article{2016AA...594A.106W,
       author = {{Weigelt}, G. and {Hofmann}, K.-H. and {Schertl}, D. and {Clementel}, N. and {Corcoran}, M.~F. and {Damineli}, A. and {de Wit}, W.-J. and {Grellmann}, R. and {Groh}, J. and {Guieu}, S. and {Gull}, T. and {Heininger}, M. and {Hillier}, D.~J. and {Hummel}, C.~A. and {Kraus}, S. and {Madura}, T. and {Mehner}, A. and {M{\'e}rand}, A. and {Millour}, F. and {Moffat}, A.~F.~J. and {Ohnaka}, K. and {Patru}, F. and {Petrov}, R.~G. and {Rengaswamy}, S. and {Richardson}, N.~D. and {Rivinius}, T. and {Sch{\"o}ller}, M. and {Teodoro}, M. and {Wittkowski}, M.},
        title = "{VLTI-AMBER velocity-resolved aperture-synthesis imaging of {\ensuremath{\eta}} Carinae with a spectral resolution of 12 000. Studies of the primary star wind and innermost wind-wind collision zone}",
      journal = {\text{Astron. Astrophys.}},
         year = 2016,
        month = oct,
       volume = {594},
          eid = {A106},
        pages = {A106},
          doi = {10.1051/0004-6361/201628832},
archivePrefix = {arXiv},
       eprint = {1610.05438},
 primaryClass = {astro-ph.SR},
       adsurl = {https://ui.adsabs.harvard.edu/abs/2016A&A...594A.106W}
}

@INPROCEEDINGS{2009AIPC.1094..792D,
       author = {{Demory}, B.-O. and {S{\'e}gransan}, D. and {Forveille}, T. and {Queloz}, D. and {Delfosse}, X. and {Perrier}, C.},
        title = "{Mass-Radius relation of low-mass stars revisited with the VLTI}",
    booktitle = {15th Cambridge Workshop on Cool Stars, Stellar Systems, and the Sun},
         year = 2009,
       editor = {{Stempels}, Eric},
       series = {American Institute of Physics Conference Series},
       volume = {1094},
        month = feb,
    publisher = {AIP},
        pages = {792-795},
          doi = {10.1063/1.3099235},
       adsurl = {https://ui.adsabs.harvard.edu/abs/2009AIPC.1094..792D}
}

@article{2004AA...418..177V,  
       author = {{van Boekel}, R. and {Waters}, L.~B.~F.~M. and {Dominik}, C. and {Dullemond}, C.~P. and {Tielens}, A.~G.~G.~M. and {de Koter}, A.},
        title = "{Spatially and spectrally resolved 10 {\ensuremath{\mu}}m emission in Herbig Ae/Be stars}",
      journal = {\text{Astron. Astrophys.}},
         year = 2004,
        month = apr,
       volume = {418},
        pages = {177-184},
          doi = {10.1051/0004-6361:20034331},
       adsurl = {https://ui.adsabs.harvard.edu/abs/2004A&A...418..177V}
}

@article{2007AA...474..837T,
       author = {{Tristram}, K.~R.~W. and {Meisenheimer}, K. and {Jaffe}, W. and {Schartmann}, M. and {Rix}, H.-W. and {Leinert}, Ch. and {Morel}, S. and {Wittkowski}, M. and {R{\"o}ttgering}, H. and {Perrin}, G. and {Lopez}, B. and {Raban}, D. and {Cotton}, W.~D. and {Graser}, U. and {Paresce}, F. and {Henning}, Th.},
        title = "{Resolving the complex structure of the dust torus in the active nucleus of the Circinus galaxy}",
      journal = {\text{Astron. Astrophys.}},
         year = 2007,
        month = nov,
       volume = {474},
       number = {3},
        pages = {837-850},
          doi = {10.1051/0004-6361:20078369},
archivePrefix = {arXiv},
       eprint = {0709.0209},
 primaryClass = {astro-ph},
       adsurl = {https://ui.adsabs.harvard.edu/abs/2007A&A...474..837T}
}
\bibliographystyle{spiebib} 

\end{document}